\documentstyle[11pt,epsf,amsbsy]{article}
\renewcommand{\vec}{\boldsymbol}
\newcounter{sectionc}\newcounter{subsectionc}\newcounter{subsubsectionc}
\renewcommand{\section}[1] {\vspace{12pt}\addtocounter{sectionc}{1}
\setcounter{subsectionc}{0}\setcounter{subsubsectionc}{0}\noindent
	{\large\bf\thesectionc. #1}\par\vspace{5pt}}
\renewcommand{\subsection}[1] {\vspace{12pt}\addtocounter{subsectionc}{1}
	\setcounter{subsubsectionc}{0}\noindent
	{\bf\thesectionc.\thesubsectionc.  #1}\par\vspace{5pt}}

\def\qed{\hbox{${\vcenter{\vbox{			
   \hrule height 0.4pt\hbox{\vrule width 0.4pt height 6pt
   \kern5pt\vrule width 0.4pt}\hrule height 0.4pt}}}$}}

\renewenvironment{thebibliography}[1]
	{\begin{list}{[\arabic{enumi}]}
        {\usecounter{enumi}\setlength{\parsep}{0pt}     
	 \setlength{\leftmargin 12.7pt}{\rightmargin 0pt} 
	 \setlength{\leftmargin 17pt}{\rightmargin 0pt}   
	 \setlength{\leftmargin 22pt}{\rightmargin 0pt}   
         \setlength{\itemsep}{0pt} \settowidth
	{\labelwidth}{#1.}\sloppy}}{\end{list}}

\begin{document}
\setcounter{page}{1}

\vspace*{1truein}
\centerline{\Large \bf{X-ray Diffraction by Time-Dependent  Deformed Crystals:}}
\vspace*{0.2truein}
\centerline{\Large \bf{Theoretical Model and Numerical Analysis  }}
\vspace*{0.2truein}\centerline{\large \bf{Svetlana Sytova}}

\begin {abstract}
\small The objective of this article is to study the behavior of electromagnetic field under X-ray diffraction by time-dependent deformed crystals. Derived system of differential equations looks like the Takagi equations in the case of non-stationary crystals. This is a system of multidimensional first-order hyperbolic equations with complex time-dependent coefficients. Efficient difference schemes based on the multicomponent modification of the alternating direction method are proposed. The stability and convergence of devised schemes are proved. Numerical results are shown for the case of an ideal crystal, a crystal heated uniformly according to a linear law and a time-varying bent crystal. Detailed numerical studies indicate the importance of consideration even small crystal changes.

\vspace*{0.1truein}
{\bf MS Classification}: 65M06 (Primary), 78A45 (Secondary)

\vspace*{0.1truein}
\end {abstract}

\normalsize
\vspace*{0.2truein}
\centerline{\large \bf{Contents}}
\vspace*{0.1truein}
\baselineskip 17pt
\begin{tabbing}
1. Introduction \hspace{12 cm} \= 2\\
2. Physical and mathematical model of X-ray dynamical diffraction by time-  \> \\
dependent deformed crystals \> 2 \\
3. Numerical analysis \> 6 \\
{\rm
\hspace{1.5cm}3.1. Difference schemes for solving hyperbolic system in two space dimensions} \> 6 \\
{\rm
\hspace{1.5cm}3.2. Stability and convergence of difference schemes} \> 7 \\
4. Results of numerical experiments \> 10 \\
{\rm
\hspace{1.5cm}4.1. Diffraction by ideal crystal} \> 10 \\
{\rm
\hspace{1.5cm}4.2. Diffraction by time-dependent heated crystal} \> 13 \\
{\rm
\hspace{1.5cm}4.3. Diffraction by time-dependent bent crystal} \> 17 \\
5. Summary \> 18 \\
6. Acknowledgements\> 19\\
7. References \> 19
\end{tabbing}
\baselineskip 13.6pt
\rm
\newpage

\section{Introduction}
\noindent

Mathematical modeling of X-ray diffraction by time-dependent deformed crystals refers to physical problems of intensive beams passing through crystals. So, a relativistic electron beam passes through the crystal target and leads to its heating and deformation. The system of differential equations describing X-ray dynamical diffraction by non-stationary  deformed crystals was obtained in ~\cite{1}. This system looks like the Takagi equations  ~\cite{2}--~\cite{3} in the case of non-stationary crystals. Up to now in many ref. (e.g.  ~\cite{4}--~\cite{7}) the theory of X-ray dynamical diffraction by stationary crystals for different deformations was developed. Proper systems of differential equations are stationary hyperbolic systems for two independent spatial variables. In  ~\cite{4}--~\cite{7} the solutions of these systems were obtained analytically for some cases of deformations.  ~\cite{8} id devoted to numerical calculation of propagation of X-rays in stationary perfect crystals and in a crystal submitted to a thermal gradient. In  ~\cite{9} the theory of time-dependent X-ray diffraction by ideal crystals was developed on the basis of Green-function formalism for some suppositions.

The exact analytical solution of the system being studied in this work is difficult if not impossible to obtain. That is why we propose difference schemes for numerical solution. To solve multidimensional hyperbolic systems it is conventional to use different componentwise splitting methods, locally one-dimensional method, alternating direction method and others. They all have one common advantage, since they allow to reduce the solving of complicated problem to solving of a system of simpler ones. But sometimes they do not give sufficient precision of approximation solution under rather wide grid spacings and low solution smoothness because the disbalancement of discrete nature causes the violation of discrete analogues of conservation laws. The alternating direction method is efficient when solving two-dimensional parabolic equations. We use the multicomponent modification of the alternating direction method  ~\cite{10} which is devoid of such imperfections. This method provides a complete approximation. It can be applied for multicomponent decomposition, does not require the operator's commutability. It can be used in solving both stationary and non-stationary problems.

Difference schemes presented allow the peculiarities of initial system solution behavior. The problem of stability and convergence of proposed difference schemes are considered. We present results of numerical experiments carried out. We compare efficiency of suggested schemes in the case of diffraction by ideal stationary crystal. Tests and results of numerical experiments  are demonstrated in the case of heated crystal. In our experiments it is assumed that the crystal was heated uniformly according to a linear law. The source of crystal heating was not specified. It may be the electron beam passing through the crystal. In  ~\cite{1} we have given the formulae which allow to determine the crystal temperature under electron beam heating. We present also results of numerical modeling of X-ray diffraction by a time-varying bent crystal. The source of crystal bending is not discussed too.

\section{Theoretical Model of X-ray Dynamical Diffraction by \\ Time-Dependent Deformed Crystals }
\noindent
We will use the physical notation  ~\cite{11}. Let a monocrystal plane be affected by some time-varying field of forces, which cause the crystal to be deformed. At the same time let a plane electromagnetic wave with frequency $\omega$ and wave vector $\vec k$ be incident on this monocrystal plane. We consider two different diffraction geometry which are depicted in Figure.1. In the case of Bragg geometry the diffracted wave leaves the crystal through the same plane that the direct wave comes in. In Laue case the diffracted wave leaves the crystal through the back plane of the crystal.
\begin{figure}[htbp]
\vspace*{13pt}
\epsfxsize = 15.5 cm
\centerline{\epsfbox{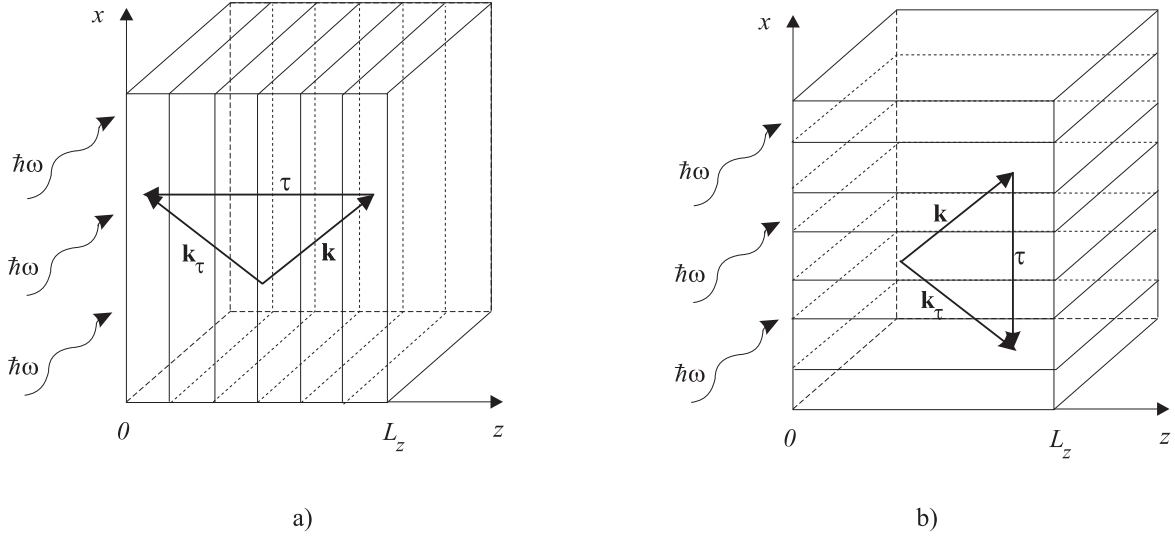}}
\vspace*{13pt}
\caption{Diffraction geometry: a) Bragg case, b) Laue case.}
\end{figure}
The electromagnetic field inside the crystal in two-wave approximation is written in the form:
\begin{displaymath}
{\bf{D}}(\vec r \it , t) = \vec D (\vec r \it ,t) \exp(i(\vec k \vec r -\omega \it t))+\vec D_{\tau } (\vec r\it ,t) \exp(i((\vec k +\vec \tau )\vec r - \omega \it t)),
\end{displaymath}
where $\vec D$ and $\vec D_{\tau}$ are the amplitudes of electromagnetic induction of direct and diffracted waves, respectively, and $ \vec \tau$ is the reciprocal lattice vector.

Let us examine a weakly distorted region in the crystal, where for the deformation vector $\vec u(\vec r, t)$ the following inequalities are correct:
\begin{displaymath}
\left| \frac{\partial \vec u}{\partial \vec r}\right | \ll 1, \qquad \left| \frac{1}{c}\frac{\partial \vec u}{\partial t}\right | \ll 1,
\end{displaymath}
where $c$ is the velocity of light.

We can write :$\vec{\tau}_d=\vec{\tau}(1-\widehat u)$. Here $\vec{\tau}_d(\vec r , t)$ is the reciprocal lattice vector in deformed crystal. $\widehat u (\vec r , t)$ is the crystal deformation tensor. $u_{ij}=1/2(\partial u_i/\partial x_j+\partial u_j/\partial x_i)$. Let us call considered system of coordinates $S$.

To obtain an expansion in series of the reciprocal lattice vector let us pass to a new system of coordinates $S'$: $\vec r_d=\vec r - \vec u(\vec r, t)$. Here in each fixed instant of time the Bravais lattice of deformed crystal is coincident with one of undistorted crystal in the system $S$. So, in the system $S'$ the crystal structure is periodic. And it is disturbed in the system $S$.

Now in $S'$ for electric susceptibility $\epsilon$ we can write:
\begin{displaymath}
\epsilon(\vec r_d; \omega)=\sum_{\vec \tau_d} \epsilon ( \vec \tau_d; \omega)exp(i \vec \tau_d \vec r_d ),
\end{displaymath}
\begin{displaymath}
\epsilon(\vec r-\vec u ;\omega )=\sum_{\vec \tau_d}\epsilon(\vec \tau_d; \omega)exp(i \vec \tau_d (\vec r -\vec u)).
\end{displaymath}
Or, finally restoring in the system $S$, we obtain:
\begin{displaymath}
\epsilon(\vec r;\omega )=\sum_{\vec \tau_d} \epsilon((1-\widehat u) \vec  \tau; \omega)exp(i \vec \tau_d \vec r),
\end{displaymath}
where $\quad \epsilon (0; \omega)=1+g_0, \quad \epsilon ((1-\widehat u)\vec \tau; \omega)=1+g_\tau (\vec r, t), \quad \epsilon (-(1-\widehat u)\vec \tau; \omega)=1+g_{-\tau} (\vec r, t).$

Let us assume that the amplitudes $\vec D$ and $\vec D_{\tau}$ are changing sufficiently slowly in the space and time:
\begin{displaymath}
\left|  {\frac{1}{k}\frac{\partial \vec D}{\partial x_i}}\right| \ll |\vec D|, \quad \left|  {\frac{1}{k}\frac{\partial \vec D_\tau }{\partial x_i}}\right| \ll |\vec D_\tau|, \quad i=1,\,2,\,3.
\end{displaymath}
\begin{displaymath}
\left|  {\frac{1}{\omega}\frac{\partial \vec D}{\partial t}}\right| \ll |\vec D|, \quad \left|  {\frac{1}{\omega}\frac{\partial \vec D_\tau }{\partial t}}\right| \ll |\vec D_\tau|.
\end{displaymath}
Then from Maxwell's equations the following system of differential equations was derived  ~\cite{1}:
\begin{eqnarray}
\frac{2i}{\omega}\frac{\partial \vec D}{\partial t}+ \frac{2i}{ k^2}\vec k  {\rm grad} \vec D + \chi_0 \vec D + \chi_{\tau} \vec D_{\tau }=0, \nonumber 
\end{eqnarray}
\begin{eqnarray}
\frac{2i}{\omega}\frac{\partial \vec D_{\tau}}{\partial t}+ \frac{2i}{k^2}\vec k_\tau  {\rm grad} \vec D_{\tau} + \chi_{-\tau} \vec D + (\chi_0-\alpha(\vec r, t)-s(\vec r, t)) \vec D_{\tau }=0,
\end{eqnarray}
where
\begin{displaymath}
\alpha(\vec r, t)=\alpha_0-\frac{2 {\vec k }_{\tau} {\rm grad} (\vec \tau \vec u )}{k^2},\quad
\alpha_0=\frac{(\tau^2+2\vec k \vec \tau )}{k^2};\quad
s(\vec r, t)=\frac {2}{\omega}\Bigl (\vec \tau \frac {\partial \vec u}{\partial t}\Bigr ); \quad  k=\frac {\omega}{c}. 
\end{displaymath}
$\chi_0$, $\chi_{\pm \tau}$ are the zero and $\pm \tau$ Fourier components of the crystal electric susceptibility.

The difference between our system (2.1) and the Takagi equations  ~\cite{3} is in the term $\alpha$, which depends on time now, and in the appearance of the term $s$.

Let us rewrite the system (2.1) in the generalized form having picked out vectors of $\sigma$-polarization from amplitudes of electromagnetic induction $\vec D$ and $\vec D_{\tau}$ and having specified three independent variables $t$, $z$, $x$. The spatial variable $y$ is a parameter. One can write a full three-dimensional system.
\begin{eqnarray}
\frac{\partial D}{\partial t} \,+A_{11} \frac{\partial D}{\partial z} +A_{12} \frac{\partial D}{\partial x} +Q_{11} D + Q_{12} D_{\tau }=0, \nonumber \\
\frac{\partial D_{\tau } }{\partial t} \,+A_{21} \frac{\partial D_{\tau } }{\partial z} +A_{22} \frac{\partial D_{\tau }} {\partial x} +Q_{21} D + Q_{22} D_{\tau } =0,
\end{eqnarray}
where
\begin{eqnarray}
A_{11}=\frac{c k_z}{k},\quad A_{12}=\frac{c k_x}{k},\quad
A_{21}=\frac{c k_{\tau z}}{k},\quad A_{22}=\frac{c k_{\tau x}}{k};
\end{eqnarray}
\begin{displaymath}
Q_{11}=-0.5 i \omega \chi_0,\quad Q_{12}=-0.5 i \omega \chi_{\tau},
\end{displaymath}
\begin{eqnarray}
Q_{21}=-0.5 i \omega \chi_{-\tau},\quad Q_{22}=-0.5 i \omega (\chi_0 - \alpha (z,x,t)- s(z,x,t)).
\end{eqnarray}

Initial and boundary conditions are written in the domain $G=\lbrace (z,x,t), 0\leq z \leq L_z, 0\leq x \leq L_x, 0 \leq t \leq T \rbrace$. In the Bragg case the boundary conditions are written as follows:
\begin{eqnarray}
D(0,x,t)&=&D_0,\nonumber\\
D_{\tau}(L_z,x,t)&=&0,\quad 0 \leq x \leq L_x,\quad t>0.
\end{eqnarray}

In Laue geometry, where the diffracted wave leaves the crystal through the crystal back plane, the boundary condition for amplitude $D_{\tau}$ should be written at $z=0$.

As is known  ~\cite{11}, the exact solution of the stationary X-ray diffraction problem has the following form:
\begin{eqnarray}
D&=&c_1\exp(i k {\delta}_1z)+c_2\exp(i k {\delta}_2z), \nonumber\\
D_{\tau}&=& c_1s_1\exp(i k {\delta}_1z)+c_2s_2\exp(i k {\delta}_2z),
\end{eqnarray}
where
${\delta}_1,{\delta}_2$ are the solutions of the dispersion equation:
\begin{displaymath}
(2\delta {\gamma}_0-{\chi}_0)(2\delta {\gamma}_1-{\alpha}_0-{\chi}_0)-{\chi}_{\tau}{\chi}_{-\tau}=0;
\end{displaymath}
\begin{displaymath}
s_i=\frac{2{\delta}_i{\gamma}_0-{\chi}_0}{{\chi}_{\tau}},\quad i=1,2;
\end{displaymath}
$\gamma_0$ and $\gamma_1$ are the cosines of the angles between $\vec k$ and $\vec k_{\tau}$, respectively, and the $z$ axis;
\begin{displaymath}
c_1=\frac{-D_0 s_2 e_2(L_z)}{s_1 e_1(L_z)-s_2 e_2(L_z)},\quad
c_2=\frac{D_0 s_1 e_1(L_z)}{s_1 e_1(L_z)-s_2 e_2(L_z)}.
\end{displaymath}
Here and below the following designations are used:
\begin{displaymath}
e_1(z)=\exp(i k {\delta}_1z), \quad
e_2(z)=\exp(i k {\delta}_2z).
\end{displaymath}

Let us impose initial conditions corresponding to the exact solution of the stationary X-ray diffraction problem in an ideal crystal:
\begin{eqnarray}
D(z,x,0)&=& c_1 e_1(z)+c_2 e_2(z), \nonumber\\
D_{\tau}(z,x,0)&=& c_1s_1 e_1(z)+c_2s_2 e_2(z),\quad 0 \leq z \leq L_z,\quad 0 \leq x \leq L_x.\nonumber
\end{eqnarray}

In the X-ray range the amplitudes (2.6) oscillate with sufficiently high frequency. For large thickness of crystal it is complicated to obtain good numerical solutions of system (2.2) with coefficients (2.4). So, let us find solution of (2.2)  for functions ${\rm D}(z,x,t)$ and ${\rm D}_{\tau}(z,x,t)$ which vary more slowly than $e_1(z)$ or $e_2(z)$:
\begin{eqnarray}
D(z,x,t)&=&{ \rm D}(z,x,t) (c_1 e_1(z)+c_2 e_2(z)), \nonumber\\
D_{\tau}(z,x,t)&=&{\rm D}_\tau (z,x,t)(c_1s_1 e_1(z)+c_2s_2 e_2(z)).
\end{eqnarray}
Then the coefficients (2.4) have to be presented by the formulae:
\begin{displaymath}
Q_{11}=-0.5 i \omega \frac{ \chi_0-2k_z /k  (c_1 {\delta}_1e_1(z)+c_2{\delta}_2e_2(z))}{c_1e_1(z)+c_2e_2(z)},
\end{displaymath}
\begin{displaymath}
Q_{12}=-0.5 i \omega\chi_{\tau}\frac{c_1s_1e_1(z)+c_2s_2e_2(z)}{c_1e_1(z)+c_2e_2(z)},
\end{displaymath}
\begin{displaymath}
Q_{21}=-0.5 i \omega \chi_{-\tau}\frac {c_1e_1(z)+c_2e_2(z)}{c_1s_1e_1(z)+c_2s_2e_2(z)},
\end{displaymath}
\begin{eqnarray}
Q_{22}=-0.5 i \omega \frac{ \chi_0 - \alpha (z,x,t)- s(z,x,t)-
2k_{\tau z}/k (c_1s_1 {\delta}_1e_1(z)+c_2s_2{\delta}_2e_2(z))}{c_1s_1e_1(z)+c_2s_2e_2(z)}.
\end{eqnarray}

The boundary conditions (2.5) take the form:
\begin{eqnarray}
{\rm D}(0,x,t)&=&1,\nonumber\\
{\rm D}_{\tau}(L_z,x,t)&=&1,\quad 0 \leq x \leq L_x,\quad t>0.
\end{eqnarray}
For this case the initial conditions have to be equal to 1 too.

\newpage
\section{Numerical Analysis}
\noindent
The original differential problem is a system of multidimensional first-order differential equations of hyperbolic type with complex-valued time-dependent coefficients. The numerical schemes employed in this work are based on the multicomponent modification of the alternating direction method. This method was originally developed in  ~\cite{10}. It turned out to be an effective way for efficient implementations of difference schemes. This method is economical and unconditionally stable without stabilizing corrections for any dimension problems of mathematical physics. It does not require spatial operator's commutability for the validity of the stability conditions. This method is efficient for operation with complex arithmetic. Its main idea is in the reduction of the initial problem to consecutive or parallel solution of weakly held subproblems in  subregions with simpler structure. That is why it allows us to perform computations on parallel computers. The main feature of this method implies that the grids for different directions can be chosen independently and for different components of approximate solution one can use proper methods.

Let introduce the Hilbert space of complex vector functions: $H=L_2(G)$. In this space the inner product and the norm are applied in the usual way:

\begin{displaymath}
(u,v)=\int\limits_{G}u(x)v(x)dx, \qquad \|u\|=(u,u)^{1/2}.
\end{displaymath}
In $H$ our system (2.2) is hyperbolic. 

\subsection{Efficient Schemes for Solving Hyperbolic System in Two Space Dimensions  }
\noindent

We use the following notation  ~\cite{12}:

\vspace*{0.1truein}
\centerline{$y_x=(y_{i+1}-y_i)/h_x$ --- right difference derivative, }

\vspace*{0.1truein}
\centerline{$y_{\overline{x}}=(y_i-y_{i-1})/h_x$  --- left  one, $\quad y_i=y(x_i);$}

\begin{displaymath}
y_t=(\widehat y-y)/h_t,\qquad \widehat y=y(t_{k+1}),\qquad y=y(t_k).
\end{displaymath}

Let us replace the domain $G$ of the continuous change of variables by the grid domain
\begin{displaymath}
G_{zxt} = \lbrace (z_i,x_j,t_k); z_i =ih_z,i=0,1,\ldots,N_1, N_1 =[L_z /h_z],x_j=jh_x,j=0,1,\ldots,N_2,
\end{displaymath}
\begin{displaymath}
N_2= [L_x/h_x],t_k=kh_t, k=0,1,\ldots,N_3, N_3=[T/h_t]\rbrace.
\end{displaymath}

The following system of difference equations approximates on $G_{zxt}$ the system (2.2) with coefficients (2.3)--(2.4):
\begin{eqnarray}
D_{t}^1+A_{11}\widehat {D}_{\overline{z}}^1+A_{12}D_x^2 +
{Q_{11} \widehat {D}^1}^*+ {Q_{12}\widehat{ D}_{\tau}^1}^*=0, \nonumber \\
D_{\tau \/t}^1+A_{21} \widehat{D} _{\tau \/z}^{1} +A_{22}D_{\tau \/x}^{2} +{Q_{21} \widehat {D}^1}^*+{\widehat {(Q_{22} D_{\tau}^1)}}^*=0,
\end{eqnarray}
\begin{eqnarray}
D_{t}^2+A_{11} \widehat{D} _{\overline{z}}^{1}+A_{12} \widehat{D} _{x}^{2}
+{Q_{11} \widehat {D}^1}^*+ {Q_{12}\widehat{ D}_{\tau}^1}^*=0,\nonumber \\
D_{\tau \/t}^{2}+A_{21} \widehat{D} _{\tau \/z}^{1} +A_{22} \widehat{D} _{\tau\/x}^{2} +{Q_{21} \widehat {D}^1}^*  + {\widehat {(Q_{22} D_{\tau}^1)}}^*=0,
\end{eqnarray}
where $ D^* =0.5(D_i +D_{i-1} ), \quad {\widehat {Q_{22}}}^*=0.5(Q_{22}(z_{i-1},x_j,t_{k+1})+Q_{22}(z_i,x_j,t_{k+1}))$ for the first equations of (3.10) and (3.11) and $ D^* =0.5(D_i +D_{i+1} ), \quad {\widehat {Q_{22}}}^*=0.5(Q_{22}(z_i,x_j,t_{k+1})+Q_{22}(z_{i+1},x_j,t_{k+1}))$ for the last ones.

For coefficients (2.3), (2.8) the system (2.2) can be approximated by the system of difference equations of the following form:
\begin{eqnarray}
D_{t}^{1}+A_{11} \widehat {D}_{\overline{z}}^{1}+A_{12}D_{x}^{2} +Q_{11}{ \widehat D}^1+ Q_{12} { \widehat D}_{\tau}^1=0, \nonumber \\
D_{\tau \/t}^{1}+A_{21} \widehat{D} _{\tau \/z}^{1} +A_{22} D_{\tau \/x}
^{2} +Q_{21}{ \widehat D}^1+{ \widehat {Q_{22} D}}_{\tau}^1 =0\nonumber  ;
\end{eqnarray}
\begin{eqnarray}
D_{t}^{2}+A_{11} \widehat{D} _{\overline{z}}^{1}+A_{12} \widehat{D} _{x}^{2}
+Q_{11}{ \widehat D}^1+Q_{12}{ \widehat D}_{\tau}^1=0 ,\nonumber \\
D_{\tau \/t}^{2}+A_{21} \widehat{D} _{\tau \/z}^{1} +A_{22} \widehat{D} _{\tau\/x}^{2} +Q_{21} { \widehat D}^1 +{ \widehat {Q_{22} D}}_{\tau}^1=0.
\end{eqnarray}

In cited schemes the directions of difference derivatives with respect to $x$ (left or right) are selected in dependence of waves directions. $D^1$, $D^2$, $D_{\tau}^1$ and $D_{\tau}^2$ are two components of approximate solutions for $D$ and $D_{\tau}$, respectively. One can choose any of these two components or its half-sum as a solution of (2.2). The boundary and initial conditions are approximated in the accurate form.

In Laue case where the diffracted wave moves on a positive direction of the z axis, for $D_{\tau}$ we should write left difference derivatives with respect to z.

The schemes (3.10)--(3.11) and (3.12) are completely consistent. The consistency clearly follows from the manner in which these schemes were constructed. On sufficiently smooth solutions they are of the first order approximation with respect to time and space. We can give the difference scheme of the second order approximation with respect to z. In this case it should be rewritten (3.10):
\begin{eqnarray}
{D_{t}^1}^*+A_{11}\widehat {D}_{\overline{z}}^1+A_{12}D_x^2 +
{Q_{11} \widehat {D}^1}^*+ {Q_{12}\widehat{ D}_{\tau}^1}^*=0, \nonumber \\
{D_{\tau \/t}^1}^*+A_{21} \widehat{D} _{\tau \/z}^{1} +A_{22}D_{\tau \/x}^{2} +{Q_{21} \widehat {D}^1}^*+{\widehat {(Q_{22} D_{\tau}^1)}}^*=0.
\end{eqnarray}
The scheme for the second component (3.11) is not changed.

One can write a scheme of the second order approximation with respect to time. But as has been shown in numerical experiments it does not lead to sensible changes in solution pattern.

For the difference schemes presented, the stability relative to initial data and also the convergence of the difference problem solution to the solution of differential problem (2.2) can be proved. This follows from the properties of the multicomponent modification of the alternating direction method  ~\cite{10}. Let us prove the corresponding Theorems.

\subsection{Stability and Convergence of Difference Schemes}
\noindent

We use the energy inequalities method  ~\cite{12}. Let rewrite the system (3.10)--(3.11) in the form:
\begin{eqnarray}
{\sf D}_t^1+\widehat \Lambda_1({\sf D}^1)+\Lambda_2({\sf D}^2)=0,\\
{\sf D}_t^2+\widehat \Lambda_1({\sf D}^1)+\widehat \Lambda_2({\sf D}^2)=0,
\end{eqnarray}
where
$$
{\sf D}=\left( \matrix{
D \cr
D_{\tau}\cr}\right), \qquad
\Lambda_1({\sf D})=\left( \matrix{
A_{11}D_{\overline{z}}+Q_{11}D^*+Q_{12}D_{\tau}^* \cr
A_{21}D_{\tau \/z}+Q_{21}D^*+(Q_{22}D_{\tau})^*\cr }\right), \qquad
\Lambda_2({\sf D})=\left( \matrix{
A_{12}D_x \cr
A_{22}D_{\tau \/x}\cr} \right).$$

Let us introduce the following notation:

\begin{displaymath}
y^\prime ={\rm Re}(y),\qquad y^{ \prime  \prime }={\rm Im}(y),\qquad \overline{y}=y^\prime -iy^{\prime  \prime} .
\end{displaymath}

We use the inner products:

\begin{displaymath}
(y,v)_{\omega}=\sum_{i=1}^{N-1}hy_iv_i,
\end{displaymath}
where $\;\omega=\lbrace x_i=ih,i=0,1,...,N, Nh=L \rbrace$ is a one-dimensional grid;
\begin{displaymath}
(Y,V)_{G_{zxt}}=(Y,V)=\sum_{i=1}^{N_1-1}\sum_{j=1}^{N_2-1}h_zh_xY_{ij}V_{ij}.
\end{displaymath}
In addition to this let us introduce the norm:
\begin{displaymath}
\|Y\|=\sqrt{(Y,Y)},\quad |Y|^2=\|Y^\prime\|^2+\|Y^{\prime \prime}\|^2 .
\end{displaymath}

\vspace*{0.1truein}
\bf{L{\sc emma}}. \it If $\; y(x_0)=0 \;$ then $\; (y,y_{\overline{x}})_{\omega}\geq 0,\;$
if $\; y(x_N)=0 \; $ then $\; (y,y_x)_{\omega}\leq 0$.

\vspace*{0.05truein}
{\bf Proof}. \rm Let us write the following transformation chain:
\begin{eqnarray}
(y,y_{\overline{x}})_{\omega}=\sum_{i=1}^{N-1} y_i(y_i-y_{i-1})=\sum_{i=1}^{N-1} (y_i^2-y_{i-1}^2-y_{i-1}(y_i-y_{i-1}))=\nonumber\\
y_{N-1}^2-y_0^2-\sum_{i=1}^{N-1} y_{i-1}(y_i-y_{i-1})=y_{N-1}^2-y_0^2-\sum_{i=0}^{N-2} y_i(y_{i+1}-y_i)=\nonumber\\
y_{N-1}^2-y_0^2+\sum_{i=0}^{N-2} y_i^2-0.5\sum_{i=0}^{N-2} (y_{i+1}+y_i)^2+0.5\sum_{i=0}^{N-2}
y_{i+1}^2+0.5\sum_{i=0}^{N-2} y_i^2=\nonumber\\
0.5y_{N-1}^2-0.5y_0^2+\sum_{i=0}^{N-2} y_i^2+\sum_{i=0}^{N-2}
y_{i+1}^2-0.5\sum_{i=0}^{N-2}(y_{i+1}+y_i)^2.\nonumber
\end{eqnarray}

The first term in the last expression is greater than or equal to 0, the second is equal to 0. Taking into consideration the inequality:
\begin{displaymath}
(y_{i+1}+y_i)^2\leq 2( y_{i+1}^2+y_i^2),
\end{displaymath}
we obtain: $\,(y,y_{\overline x})_{\omega}\geq 0$. Second Lemma's inequality is proved similarly. $\Box$

\vspace*{0.1truein}
\bf {T{\sc eorem} 3.1}. \it The difference scheme (3.10)--(3.11) is unconditionally stable  relative to the initial data. For its solution the following estimates hold:
\begin{eqnarray}
|{\widehat{\sf D}}^i|^2 \leq  M \left( |{{\sf D}^i(t_0)}|^2+
|\Lambda_1({\sf D}^1(t_0))+\Lambda_2({\sf D}^2(t_0))|^2 \right),
\end{eqnarray}
where $M$ is a bounded positive constant independent of grid spacings, $i=1,2$.

\vspace*{0.05truein}
{\bf Proof}. \rm Multiply (3.14) by $\overline {\left ( \Lambda_1({\sf D}^1)\right )_t}$, (3.15) by $\overline {\left ( \Lambda_2({\sf D}^2)\right )_t}$  and sum :
\begin{eqnarray}
\left ({({\sf D}_t^1)}^\prime ,(\Lambda_1({\sf D}^1))_t^\prime \right )+\left (({\sf D}_t^1)^{\prime \prime} ,(\Lambda_1({\sf D}^1))_t^{\prime \prime} \right )+\nonumber\\
\left (({\sf D}_t^2)^\prime ,(\Lambda_2({\sf D}^2))_t^\prime \right )+\left (({\sf D}_t^2)^{\prime \prime} ,(\Lambda_2({\sf D}^2))_t^{\prime \prime} \right )+\nonumber\\
\left ((\widehat\Lambda_1({{\sf D}}^1))^\prime ,(\Lambda_1({\sf D}^1))_t^\prime \right )+
\left ((\widehat\Lambda_1({{\sf D}}^1))^{\prime \prime} ,(\Lambda_1({\sf D}^1))_t^{\prime \prime} \right )+\nonumber\\
\left ((\Lambda_2({\sf D}^2))^\prime ,(\Lambda_1({\sf D}^1))_t^\prime \right )+
\left ((\Lambda_2({\sf D}^2))^{\prime  \prime} ,(\Lambda_1({\sf D}^1))_t^{\prime  \prime} \right )+\nonumber\\
\left ((\widehat\Lambda_1({{\sf D}}^1))^\prime ,(\Lambda_2({\sf D}^2))_t^\prime \right )+
\left ((\widehat\Lambda_1({{\sf D}}^1))^{\prime \prime} ,(\Lambda_2({\sf D}^2))_t^{\prime \prime} \right )+\nonumber\\
\left ((\widehat\Lambda_2({{\sf D}}^2))^\prime ,(\Lambda_2({\sf D}^2))_t^\prime \right )+
\left ((\widehat\Lambda_2({{\sf D}}^2))^{\prime \prime} ,(\Lambda_2({\sf D}^2))_t^{\prime \prime} \right )=0.
\end{eqnarray}

Let us multiply (3.17) by $h_t$ and take into account the form of time derivatives. Then we obtain:
\begin{eqnarray}
\Phi({\sf D})+0.5\left (|\widehat\Lambda_1({{\sf D}}^1)+\widehat\Lambda_2({{\sf D}}^2)|^2+
|\Lambda_1({\sf D}^1)+\Lambda_2({\sf D}^2)|^2\right )+\nonumber\\
0.5\left (|\widehat\Lambda_1({\sf D}^1)|^2+|\widehat\Lambda_2({\sf D}^2)|^2+
|\Lambda_1({\sf D}^1)|^2+|\Lambda_2({\sf D}^2)|^2\right )-\nonumber\\
\left ((\widehat\Lambda_1({{\sf D}}^1))^\prime,(\Lambda_1({\sf D}^1))^\prime\right )-
\left ((\widehat\Lambda_2({{\sf D}}^2))^{\prime },(\Lambda_2({\sf D}^2))^{\prime}\right )-\nonumber\\
\left ((\widehat\Lambda_1({{\sf D}}^1))^{\prime \prime},(\Lambda_1({\sf D}^1))^{\prime\prime}\right )-
\left ((\widehat\Lambda_2({{\sf D}}^2))^{\prime \prime},(\Lambda_2({\sf D}^2))^{\prime \prime}\right )=0,
\end{eqnarray}
where
\begin{eqnarray}
\Phi({\sf D})=h_t\left ((D_t^1)^\prime,(\Lambda_1({\sf D}^1))_t^\prime\right )+
h_t\left ((D_t^1)^{\prime\prime},(\Lambda_1({\sf D}^1))_t^{\prime\prime}\right )+
\nonumber\\
h_t\left ((D_t^2)^\prime,(\Lambda_2({\sf D}^2))_t^\prime\right )+
h_t\left ((D_t^2)^{\prime\prime},(\Lambda_2({\sf D}^2))_t^{\prime\prime}\right ).
\end{eqnarray}

Let us re-arrange (3.18):
\begin{eqnarray}
\Phi({\sf D})+0.5|\widehat \Lambda_1({\sf D}^1)+
\widehat\Lambda_2({\sf D}^2)|^2-0.5|\Lambda_1({\sf D}^1)+\Lambda_2({\sf D}^2)|^2+\nonumber\\
0.5h_t^2|(\Lambda_1({\sf D}^1))_t|^2+0.5h_t^2|(\Lambda_2({\sf D}^2))_t|^2=0.\nonumber
\end{eqnarray}
Or, more properly,
\begin{eqnarray}
\Phi({\sf D})+0.5|\widehat\Lambda_1({\sf D}^1)+\widehat\Lambda_2({\sf D}^2)|^2 \leq 0.5|\Lambda_1({\sf D}^1)+\Lambda_2({\sf D}^2)|^2.
\end{eqnarray}

If we prove that $\Phi({\sf D}) \geq 0$ then the estimate (3.16) will be obtained for $i=2$. We consider the first term in (3.19)
\begin{eqnarray}
\left ((D_t^1)^\prime,(\Lambda_1({\sf D}^1)_t^\prime \right )+\left ((D_t^1)^{\prime \prime},(\Lambda_1({\sf D}^1)_t^{\prime \prime}\right ) = \nonumber\\
\left ((D_t^1)^\prime,(A_{11}D_{\overline {z}}^1)_t^\prime \right )-
\left ((D_t^1)^\prime,Q_{11}(D^{1*})_t^{\prime \prime} \right )-\left ((D_t^1)^\prime,Q_{12}(D_{\tau}^{1*})_t^{\prime \prime} \right )+ \nonumber\\
\left ((D_{\tau\/t}^1)^\prime,(A_{21}D_{\tau\/z}^1)_t^\prime \right )-
\left ((D_{\tau\/t}^1)^\prime,Q_{21}(D^{1*})_t^{\prime\prime} \right )-\left ((D_{\tau\/t}^1)^\prime,(Q_{22}{D_{\tau}^{1*}})_t^{\prime \prime} \right ) + \nonumber\\
\left((D_t^1)^{\prime\prime},(A_{11}D_{\overline {z}}^1)_t^{\prime\prime} \right )+
\left((D_t^1)^{\prime\prime},Q_{11}(D^{1*})_t^{\prime} \right )+\left((D_t^1)^{\prime\prime},Q_{12}(D_{\tau}^{1*})_t^{\prime} \right )+ \nonumber\\
\left((D_{\tau\/t}^1)^{\prime\prime},(A_{21}D_{\tau\/z}^1)_t^{\prime\prime} \right )+
\left((D_{\tau\/t}^1)^{\prime\prime},Q_{21}(D^{1*})_t^{\prime} \right )+\left((D_{\tau\/t}^1)^{\prime\prime},(Q_{22}{D_{\tau}^{1*}})_t^{\prime} \right ) \geq 0,
\end{eqnarray}
as since from Lemma we have:
\begin{displaymath}
\left((D_t^1)^\prime,(A_{11}D_{\overline {z}}^1)_t^\prime \right)\geq 0, \quad
\left((D_t^1)^{\prime\prime},(A_{11}D_{\overline {z}}^1)_t^{\prime\prime} \right)\geq 0,
\quad A_{11}=(A_{11})^\prime >0;
\end{displaymath}
\begin{displaymath}
\left((D_{\tau\/t}^1)^\prime,(A_{21}D_{\tau\/z}^1)_t^\prime \right)\geq 0, \quad
\left((D_{\tau\/t}^1)^{\prime\prime},(A_{21}D_{\tau\/z}^1)_t^{\prime\prime} \right)\geq 0,
\quad A_{21}=(A_{21})^\prime <0.
\end{displaymath}
In (3.21) we took into account that the coefficients $Q$ from (2.4) were pure imaginary, $Q_{12}=Q_{21}$. Also we have written out corresponding inner products.  

In the same fashion as Lemma we obtain for the second term in (3.19):
\begin{eqnarray}
\left ((D_t^2)^\prime ,(\Lambda_2(\widehat{{\sf D}}^2)_t^\prime  \right )+\left ((D_t^2)^{\prime \prime},(\Lambda_2(\widehat{{\sf D}}^2)_t^{\prime \prime} \right ) \geq 0.
\end{eqnarray}

So, expressions (3.21) and (3.22) mean that operators $\Lambda_1$ and $\Lambda_2$ are positive definite.

If we repeatedly apply the expression (3.20) to the right and if we take into account that
\begin{eqnarray}
{\sf D}^2_t=-\widehat\Lambda_1({\sf D}^1)-\widehat\Lambda_2({\sf D}^2),
\end{eqnarray}
we have the estimate (3.16) for $i=2$.

Now let us obtain the estimate (3.16) for $i=1$. Multiply $\widehat {\sf D}^1$  by $\overline {\widehat {\sf D}^1}$ and take into account the expression which follows from our designations:
\begin{displaymath}
\widehat{\sf D}^1={\sf D}^1-h_t\left (\widehat\Lambda_1({\sf D}^1)+\Lambda_2({\sf D}^2)\right ).
\end{displaymath}
Using $\epsilon$-inequality  ~\cite{12} $\; |(u,v)|\leq \epsilon \|u\|^2+1/(4\epsilon)\|v\|^2$, $(\epsilon>0)$, we have :
\begin{eqnarray}
|{\widehat{\sf D}}^1|^2=\left ({\sf D}^1-h_t\left(\widehat\Lambda_1({\sf D}^1)+\Lambda_2({\sf D}^2)\right),\overline{{\sf D}^1-h_t\left(\widehat\Lambda_1({\sf D}^1)+\Lambda_2({\sf D}^2)\right)}\right )=\nonumber\\
|{\sf D}^1|^2+h_t^2|\widehat\Lambda_1({\sf D}^1)+\Lambda_2({\sf D}^2)|^2-2h_t\left(({\sf D}^1)^\prime,(\widehat\Lambda_1({\sf D}^1)+\Lambda_2({\sf D}^2))^\prime\right)\nonumber\\
-2h_t\left(({\sf D}^1)^{\prime\prime},(\widehat\Lambda_1({\sf D}^1)+\Lambda_2({\sf D}^2))^{\prime\prime}\right)\leq
M_1|{\sf D}^1|^2+M_2|\widehat\Lambda_1({\sf D}^1)+\Lambda_2({\sf D}^2)|^2.\nonumber
\end{eqnarray}
Let us consider the second term in previous inequality:
\begin{eqnarray}
|\widehat\Lambda_1({\sf D}^1)+\Lambda_2({\sf D}^2)|^2=|\widehat\Lambda_1({\sf D}^1)+\widehat\Lambda_2({\sf D}^2)-h_t(\Lambda_2({\sf D}^2))_t|^2=\nonumber\\
|\widehat\Lambda_1({\sf D}^1)+\widehat\Lambda_2({\sf D}^2)|^2+
h_t^2|(\Lambda_2({\sf D}^2))_t|^2-
2h_t\left(( \widehat\Lambda_1({\sf D}^1)+\widehat\Lambda_2({\sf D}^2))^\prime,( \Lambda_2({\sf D}^2))^\prime _t\right)-\nonumber\\
2h_t\left( ( \widehat\Lambda_1({\sf D}^1)+\widehat\Lambda_2({\sf D}^2))^{\prime \prime},( \Lambda_2({\sf D}^2))^{\prime \prime} _t\right)\leq
|\widehat\Lambda_1({\sf D}^1)+\widehat\Lambda_2({\sf D}^2)|^2.\nonumber
\end{eqnarray}
This was obtained by taking into account (3.23) and (3.22). Finally we have:
\begin{displaymath}
|{\widehat{\sf D}}^1|^2\leq M_3\left(|{\sf D}^1|^2+|\Lambda_1({\sf D}^1)+\Lambda_2({\sf D}^2)|^2 \right).
\end{displaymath}
Repeated application of this inequality to the right yields the estimate (3.16) for $i=1$. $\Box$

\vspace*{0.1truein}
We denote the discretization error as
$\,Z^i={\sf D}^i-{\sf D}, \, i=1,2,\,$
where {\sf D} is the exact solution of initial differential problem.

\vspace*{0.1truein}
\bf{T{\sc eorem} 3.2}. \it Let the differential problem (2.2)--(2.5) have a unique solution. Then the solution of the difference problem (3.10)--(3.11) converges to the solution of the initial differential problem as $h_t, h_z, h_x \to 0$. The discretization error may be written as
\begin{displaymath}
|Z^i|\leq O(h_t+h_z+h_x).
\end{displaymath}

\vspace*{0.05truein}
\bf{Proof} \rm follows immediately from consistency of the scheme (3.10)--(3.11), Theorem 3.1 and Lax's Equivalence Theorem  ~\cite{13}. $\Box$

\vspace*{0.1truein}
The stability and convergence of schemes (3.12), (3.13)--(3.11) can be proved in an analogous way.

\section{Results of Numerical Experiments}
\noindent

\subsection{Diffraction by Ideal Crystal  }
\noindent

The problem of studying of electromagnetic fields under X-ray diffraction inside the crystal target is constituent of the problem of modeling intensive beams passing through crystals, X-ray free electron laser and others. Therefore let analyze the operation of schemes (3.10)--(3.11) and (3.12) in the case of ideal absorbing crystals.

Figure 2--5 display results of numerical experiments in the crystal of LiH. This crystal was chosen because of small absorption. The design parameters were the following. The frequency $\omega$ was equal to $1.7 \cdot 10^{19} {\rm sec}^{-1}$. The diffraction plane indexes were (2, 2, 0). The Bragg angle was equal to 0.39. The angle between the direct wave vector and the $z$ axis was equal to 0.83. This case corresponds to the total internal reflection region. We compare the numerical results obtained with the analytical solution (2.6). On our figures this solution is depicted by red curves. So, Figure 2 presents numerical results of the scheme (3.10)--(3.11) for $N_1=100$ and $N_1=200$ respectively. From these plots we notice that the grid dimension $N_1=100$ gives good agreement with analytical results and $N_1=200$ gives ideal agreement. But for greater  thickness of $L_z=0.3$ cm only $N_1=200$ gives a more or less acceptable fit, as is obvious from Figure 3. 

Thus, for large thickness of crystal we will use the scheme (3.12) with coefficients (2.8). It is evident that the amplitudes ${\rm D}$ and ${\rm D}_\tau$ from (2.7) should be equal to 1 after installation of a stationary regime in the system in the case of an ideal crystal. Figure 4 demonstrates the distinction between numerical results for $N_1=20$ (curve 1), $N_1=50$ (curve 2) and 1. When $N_1=100$ the agreement is ideal. For the amplitudes of diffracted waves we can show similar figures. Figure 4 b) presents numerical and analytical solutions for the amplitudes of direct wave when $N_1=50$.
\begin{figure}[tbp]
\vspace*{13pt}
\epsfxsize = 15.5 cm
\centerline{\epsfbox{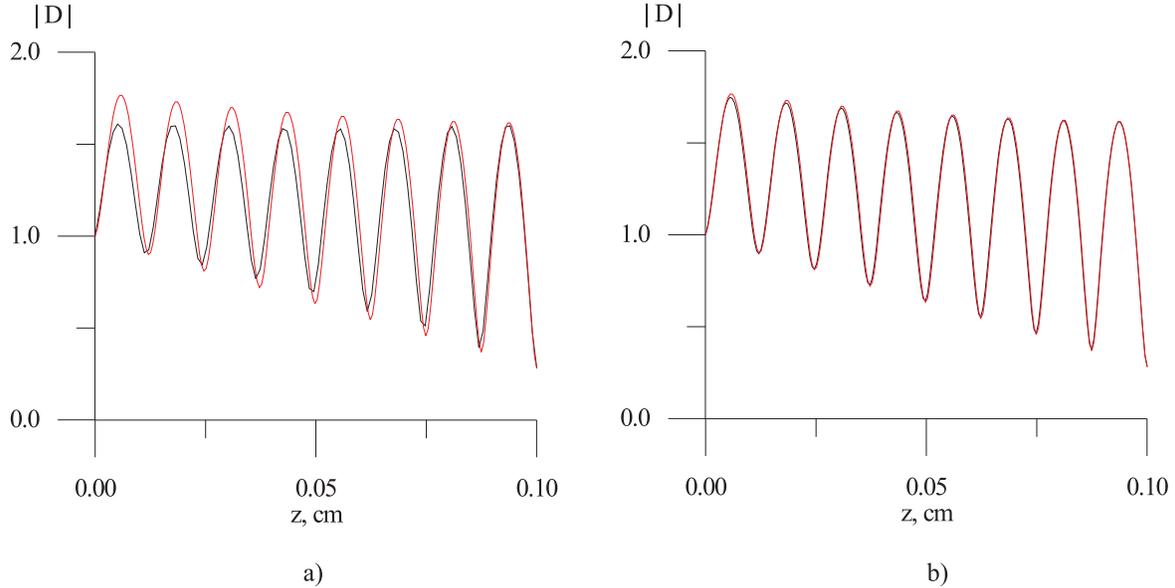}}
\vspace*{13pt}
\caption{ Amplitudes of direct wave for $L=0.1$ cm, a) $N_1=100$, b) $N_1=200$. }
\end{figure}
\begin{figure}[htbp]
\vspace*{13pt}
\epsfxsize = 15.5 cm
\centerline{\epsfbox{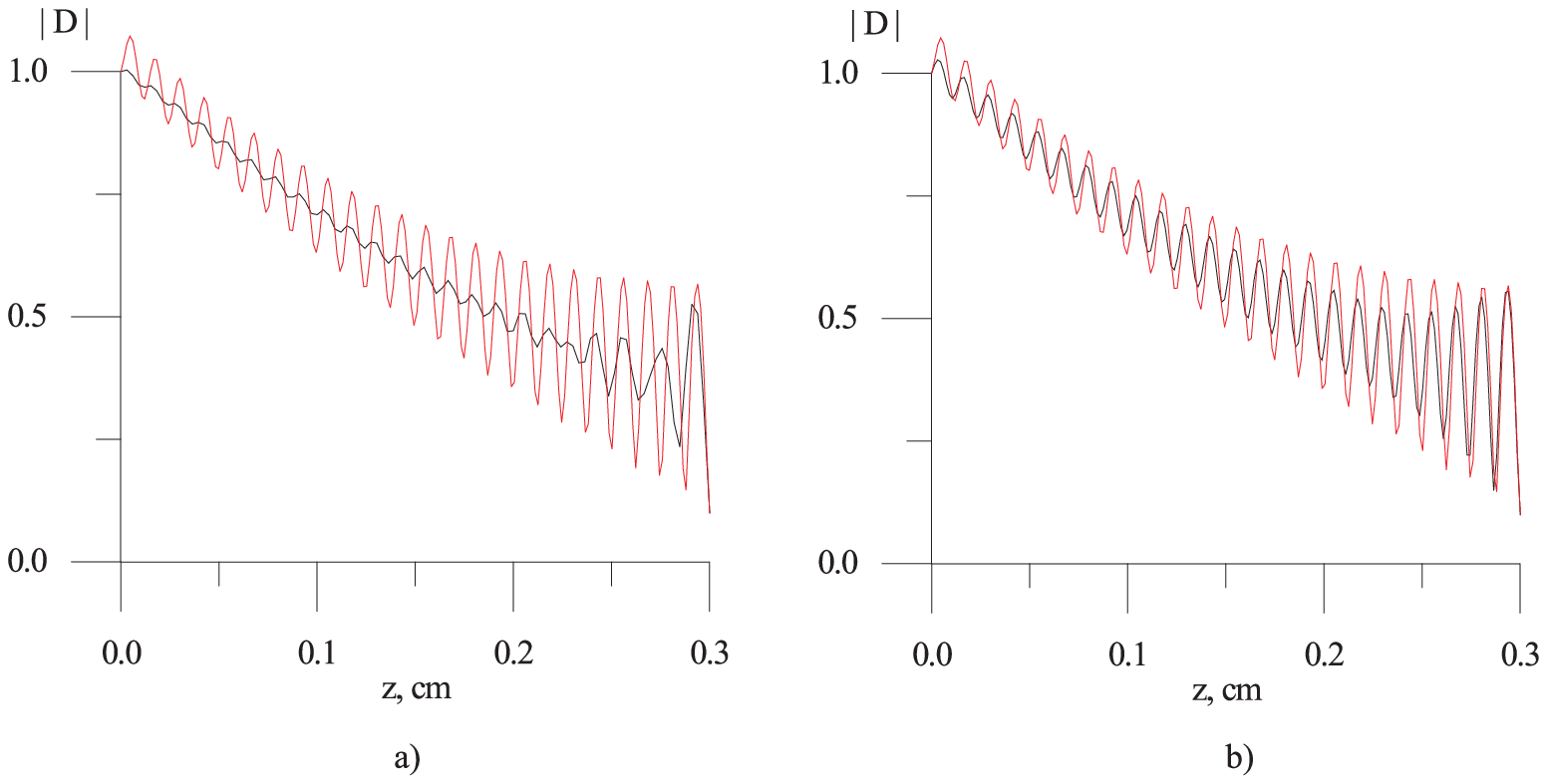}}
\vspace*{13pt}
\caption{ Amplitudes of direct wave for $L=0.3$ cm, a) $N_1=100$, b) $N_1=200$. }
\end{figure}
\begin{figure}[htbp]
\vspace*{13pt}
\epsfysize = 7.0 cm
\centerline{\epsfbox{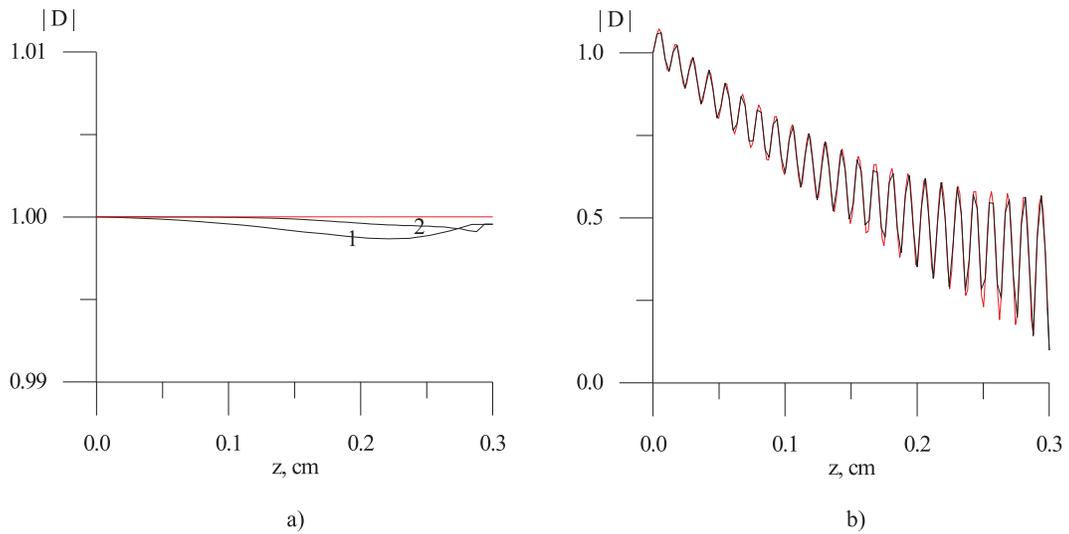}}
\vspace*{13pt}
\caption{ a) Reduced amplitudes of direct wave for $L=0.3$ cm, 1 --- $N_1=20$, 2 --- $N_1=50$ by scheme (3.12) with coefficients (2.8).
b) Amplitudes of direct wave for $L=0.3$ cm, $N_1=50$ (numerical and analytical solutions).  }
\end{figure}

Let us clear up how the scheme (3.12) functions with coefficients (2.3)-(2.4). Figure 5 gives an idea that in the case of oscillations of amplitudes this scheme operates badly even for small thickness of crystal.
\begin{figure}[htbp]
\vspace*{13pt}
\epsfxsize = 15.5 cm
\centerline{\epsfbox{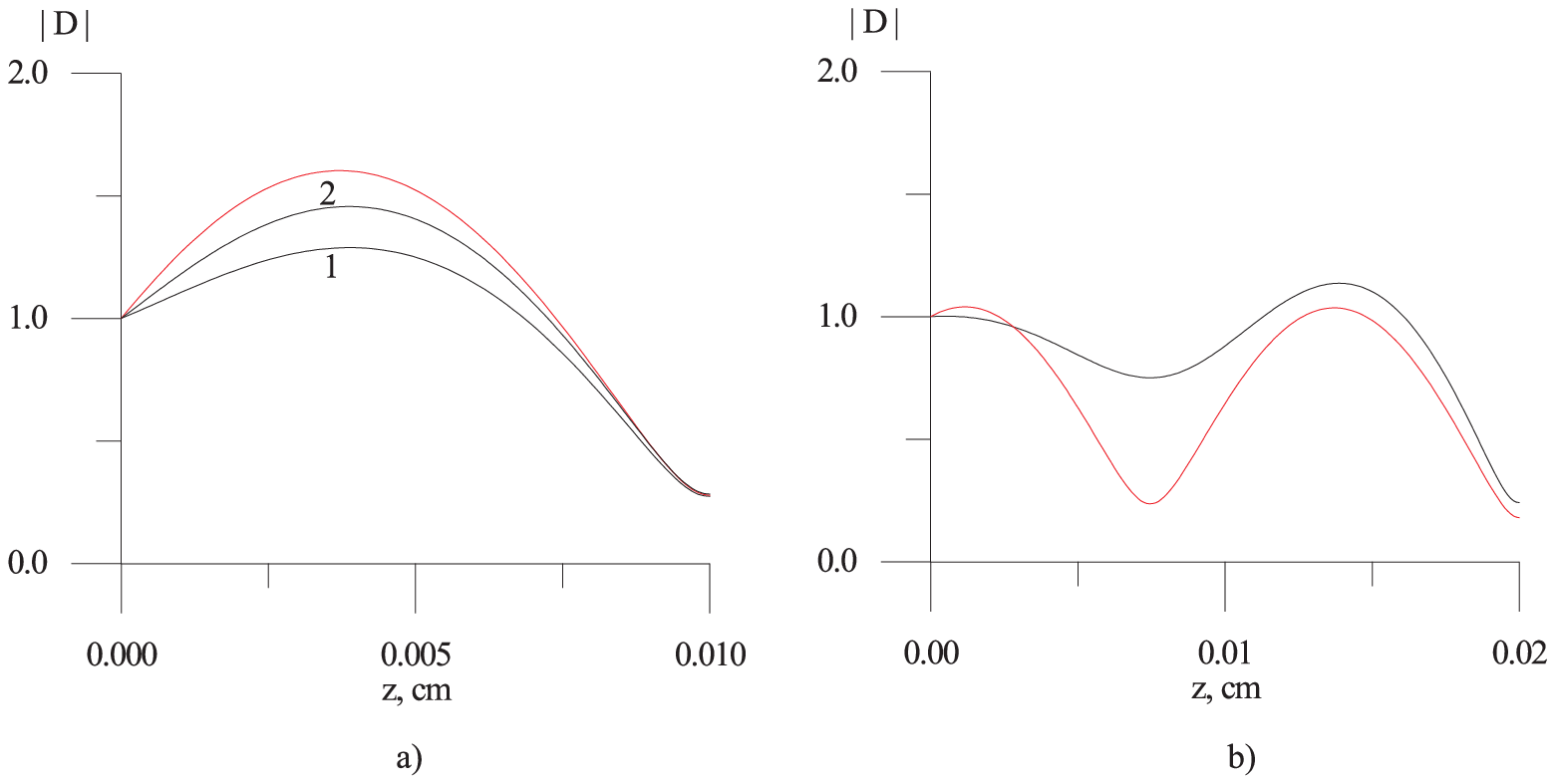}}
\vspace*{13pt}
\caption{ Amplitudes of direct wave for a) $L=0.01$ cm, 1 --- $N_1=100$, 2 --- $N_1=200$ b) $L=0.02$ cm,  1 --- $N_1=100$, 2 --- $N_1=200$ by scheme (3.12) with coefficients (2.4).}
\end{figure}

Let us show numerical results for the crystal of Si with small thickness $L_z=0.005$ cm. They are more visual because the absorption coefficient of Si is large. We have used the following geometry parameters: the diffraction plane indexes are (2, 2, 0), $\omega=6.9 \cdot 10^{18} {\rm sec}^{-1}$. The Bragg diffraction case was modeled with the Bragg angle equal to $\pi /4$.  Figure 6 depicts curves of amplitudes of direct and diffracted waves in comparison with analytical solution in the ideal crystal (2.6) (red curves). Figure 7, a) represents comparison between results of schemes (3.10)--(3.11) and (3.12) with coefficients (2.4). As stated above, the simplest scheme does not work well in our conditions. Figure 7, b) shows the behavior of two components $D^1$ and $D^2$ of numerical solution.  As may be seen from this figure, both of two components converge well to the analytical solution. That supports once again our statements given above.
\begin{figure}[htbp]
\vspace*{13pt}
\epsfxsize = 15.5 cm
\centerline{\epsfbox{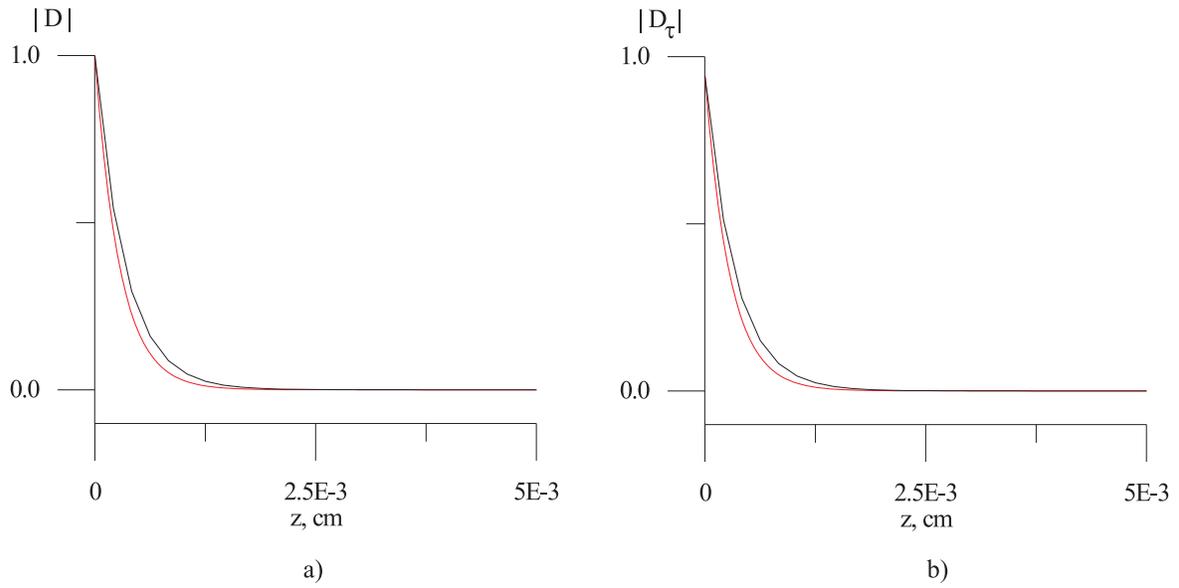}}
\vspace*{13pt}
\caption{ Amplitudes of a) direct wave and b) diffracted wave in the crystal of Si by scheme (3.10)--(3.11) with coefficients (2.4).}
\end{figure}
\begin{figure}[htbp]
\vspace*{13pt}
\epsfxsize = 15.5 cm
\centerline{\epsfbox{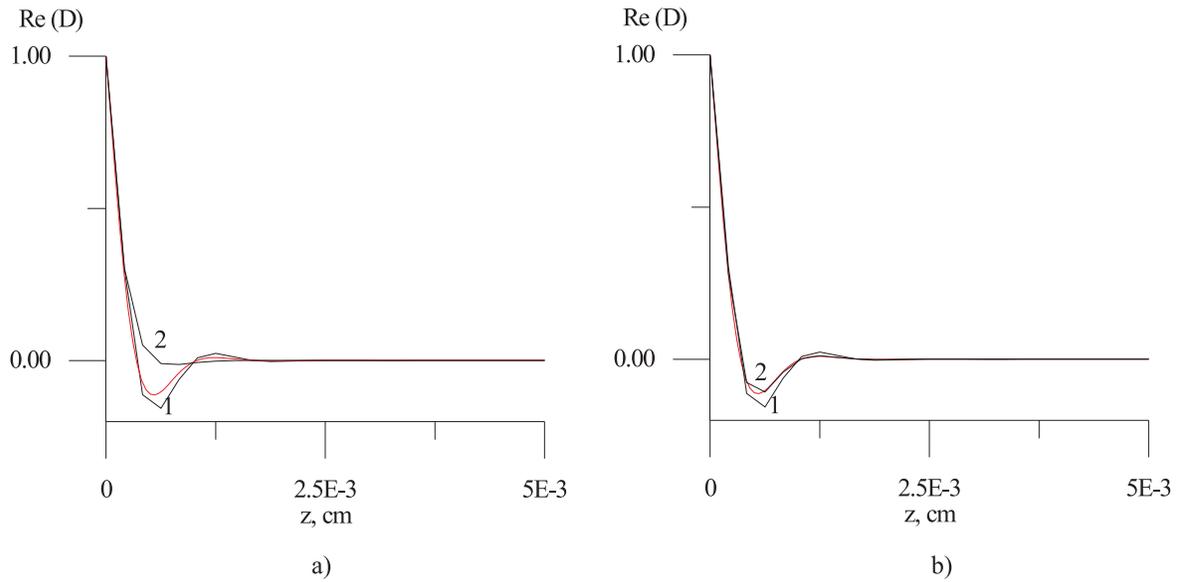}}
\vspace*{13pt}
\caption{ Amplitudes of direct wave a) for the schemes (3.10)--(3.11) (curve 1) and (3.12) (curve 2); b) two components $D^1$ (curve 1) and $D^2$ (curve 2) of numerical solution (3.10)--(3.11).}
\end{figure}

\subsection{Diffraction by Time-Dependent Heated Crystal  }
\noindent

Let us compare obtained numerical results for the crystal heated to the temperature ${\rm T}=T-T_0$ K with analytical solutions of the stationary linear X-ray diffraction problem in crystal heated to ${\rm T}$ K. $T_0$ is an initial temperature of the crystal. The source of crystal heating and deformation was not specified. It was supposed that the crystal was heated uniformly according to the linear law: $T(t)=T_0+at$, where $a$ is the rate of heating. The data for the stationary linear X-ray diffraction problem $(\chi_0, \chi_{\pm \tau}({\rm T}))$ were obtained from the program  ~\cite{14}. Let emphasize that only the values $\chi_0$ and $ \chi_{\pm \tau}(T_0)$ in numerical calculations by (3.10)--(3.11) are needed. They can be taken from reference books or from the program  ~\cite{14}. While computing the values of $\alpha$ and $s$ from the coefficient $Q_{22}$ are recalculated depending on the variation of the deformation vector $\vec u (z, x, t)$.

\begin{figure}[tbp]
\epsfysize = 7 cm
\centerline{\epsfbox{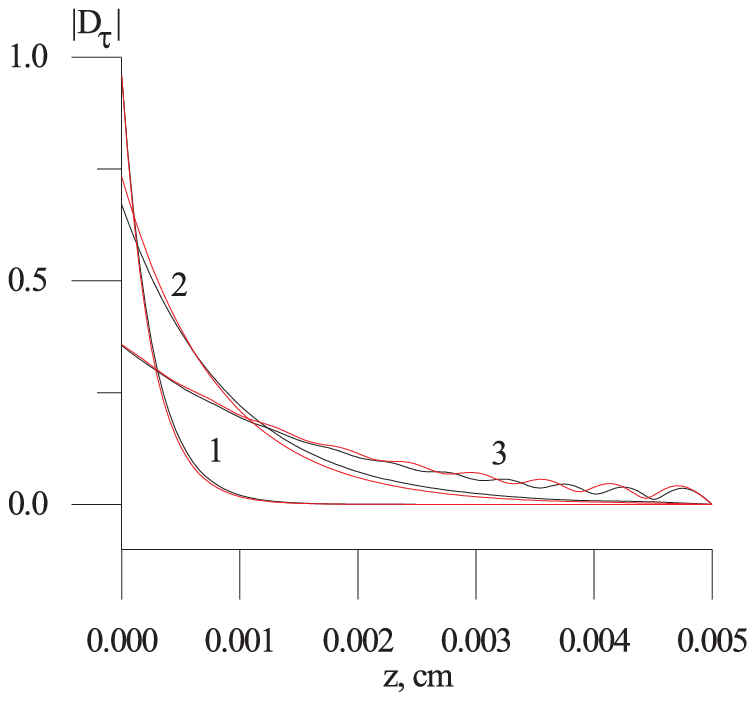}}
\caption{ Amplitudes of diffracted wave when crystal of Si heating to ${\rm T}=10$ K (curves 1), ${\rm T}=15$ K (curves 2), ${\rm T}=20$ K (curves 3). } 
\end{figure}

\begin{figure}[tbp]
\epsfxsize = 10.5 cm
\epsfysize = 11 cm
\centerline{\epsfbox{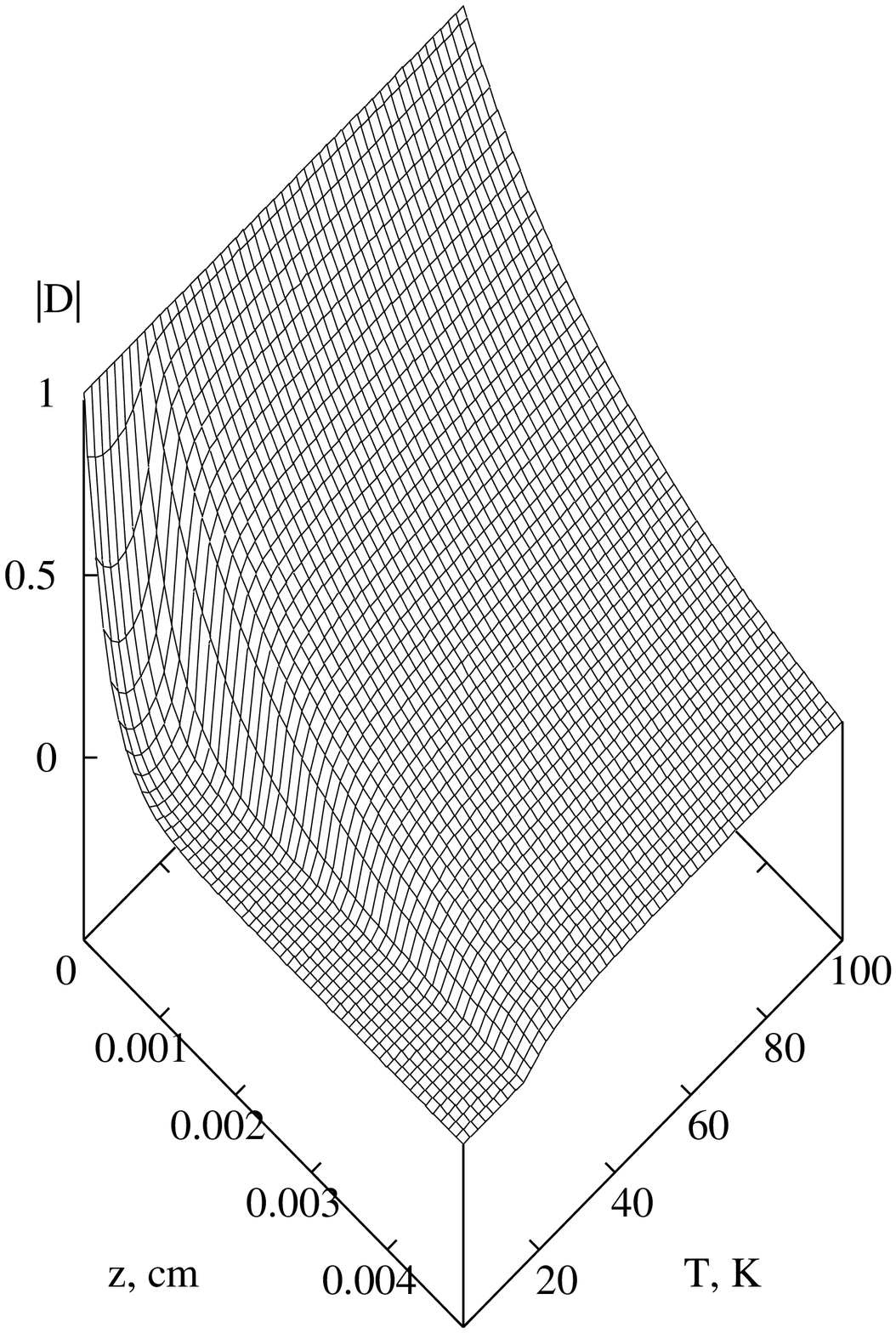}}
\caption{ Evolution of direct wave amplitude under crystal heating to 100 K.}
\end{figure}

\begin{figure}[tbp]
\epsfxsize = 10.5 cm
\epsfysize = 11 cm
\centerline{\epsfbox{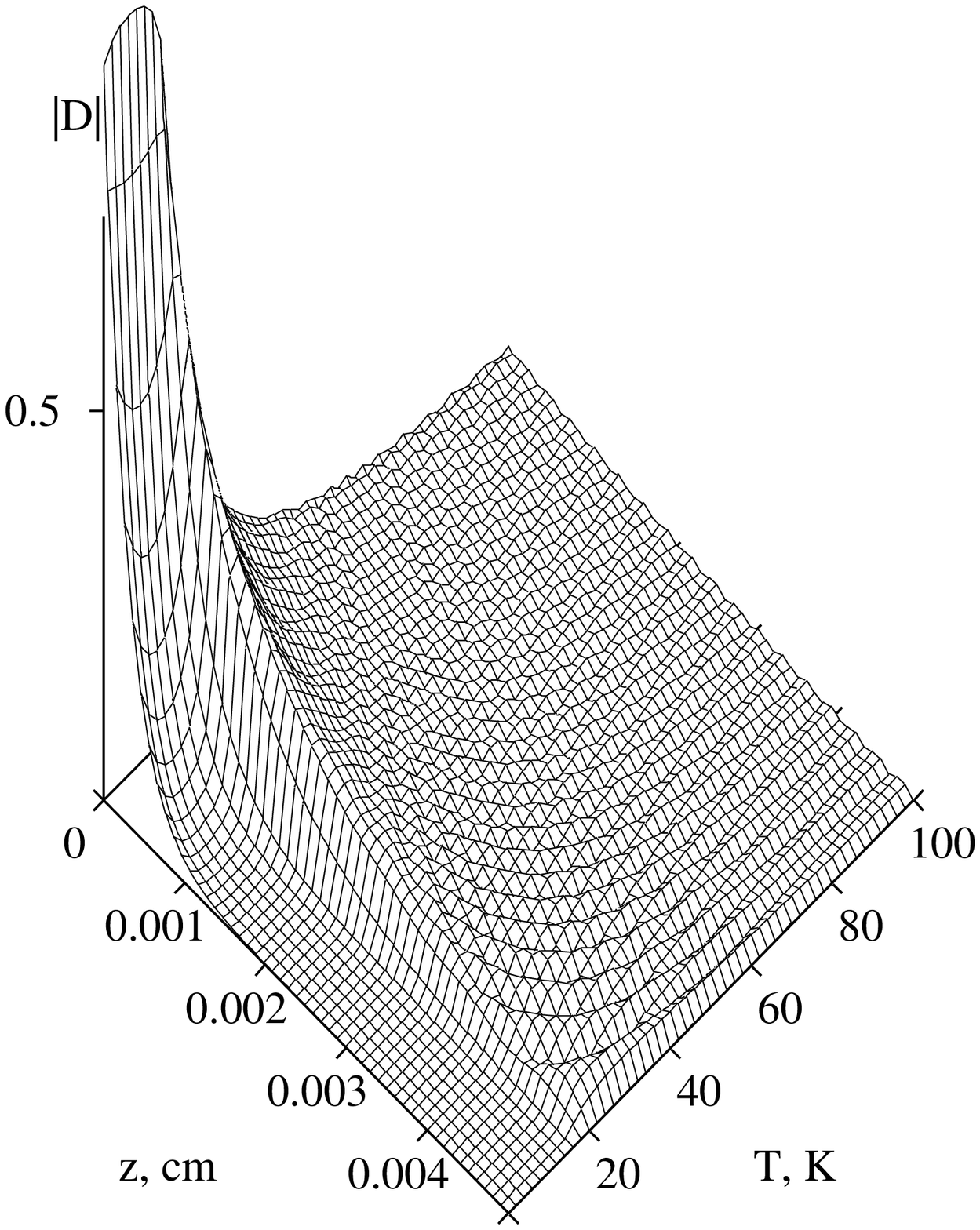}}
\caption{ Evolution of diffracted wave amplitude under crystal heating to 100 K.}
\end{figure}

We demonstrate numerical results for the crystal of Si. Figure 8  depicts curves of amplitudes of diffracted wave in the crystal of Si for the above parameters and ${\rm T}=10$ K (curves 1), ${\rm T}=15$ K (curves 2), ${\rm T}=20$ K (curves 3), respectively. The initial temperature $T_0$ was equal to 293 K. The heating rate $a$ was equal to $5 \cdot 10^{10}$ K/sec. The curves of each of pairs of curves correspond to the numerical solution of the X-ray dynamical diffraction problem in time-varying heated crystal and to the analytical solution of the stationary linear X-ray diffraction problem in the heated deformed crystal. Figure 9 and Figure 10 show evolution of direct and diffracted wave amplitudes under crystal heating to 100 K, respectively. Apparently, up to ${\rm T}=16$ K the modulus of the amplitude of the diffracted wave coming out the crystal at $z=0$ decreases abruptly. This can be explain by the fact that under heating the parameter $\alpha$ of deviation from the exact Bragg condition increases and diffraction disrupts. Such an analysis of results demonstrates that proposed mathematical model and effective numerical algorithm allow to obtain distributions of electromagnetic waves amplitudes in non-stationary crystals with sufficient precision.

\subsection{Diffraction by Time-dependent  Bent Crystal  }
\noindent

Let us examine the following model of bent in time crystal. As before, we do not specify the nature of bending (mechanic, temperature or other). We suppose that the crystal is bent according to the law
\begin{eqnarray}
u_z(x,t)=atx^2
\end{eqnarray}
(see Figure 11), where $a$ is the rate of bending. 
\begin{figure}[tbp]
\vspace*{13pt}
\epsfysize = 6. cm
\centerline{\epsfbox{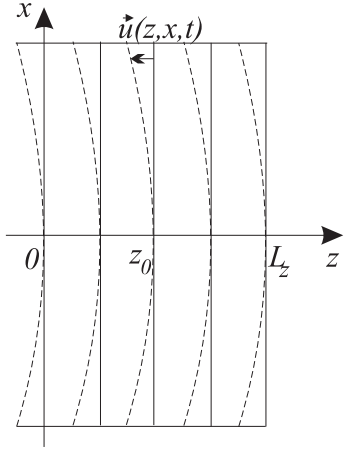}}
\vspace*{13pt}
\caption{Model of crystal bending.}
\end{figure}
(4.24) was obtained from the following assumptions. The crystal plane formula at the point $z_0$ is:
\begin{eqnarray}
z=z_0.
\end{eqnarray}
The parabola formula at the point $z_0$ has the form:
\begin{eqnarray}
atx^2=z-z_0.
\end{eqnarray}
The difference between (4.25) and (4.24) is the $z$ component of the deformation vector $\vec u(z, x, t)$. The component $u_x(z,x,t)$ can be derived from the formula:
\begin{displaymath}
u_x(z,x,t)=x_0-x,
\end{displaymath}
where $x_0$ is found from the curve distance formula
\begin{displaymath}
\int\limits_0^{x_0}\sqrt{1+4a^2t^2{\xi}^2}d{\xi}=x,
\end{displaymath}
or:
\begin{displaymath}
\frac{x_0}{2}\sqrt{1+4a^2t^2x_0^2}+\frac{1}{4at}\ln \Bigl |2atx_0+\sqrt{1+4a^2t^2x_0^2}\Bigr |=x.
\end{displaymath}
But the estimations show that when the magnitude of $u_z$ is not large (of the order $10^{-6}$), the magnitude of  $u_x$ is
of the order $10^{-8}$ and can be neglected.

We have realized numerical experiments to find out how crystal bending affects the diffraction pattern. We consider the crystal of Si with the rate of bending $a=-2.5  \cdot  10^6 {\rm (cm \, sec)}^{-1}$. Figure 12 illustrates evolution of diffraction at the point $x=0.1$ cm. The pattern of distribution of electromagnetic field is similar to one under crystal heating. When moving to the central point $x=0$ of the crystal the magnitude of bending becomes smaller. The diffraction pattern should be less changed. This fact was confirmed during numerical experiments.

\begin{figure}[tbp]
\epsfxsize = 10.5 cm
\epsfysize = 10 cm
\centerline{\epsfbox{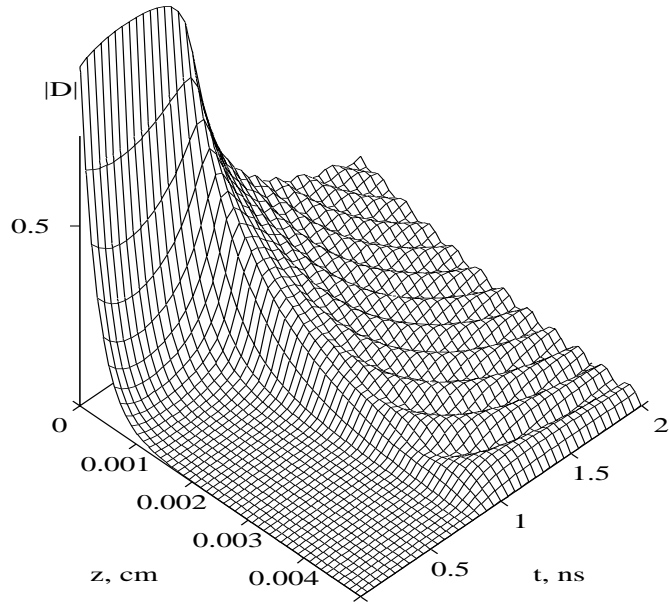}}
\caption{ Amplitudes of diffracted  wave under crystal bending at $x=0.1$ cm.}
\end{figure}

We considered simple models of crystal heating and bending.  More accurate ones can be taken, for example, from  ~\cite{15}. In ~\cite{4}-\cite{7} a general dynamical theory of X-ray diffraction from a homogeneously bent stationary crystals was developed. But their analytical formulae are complicated enough. That is why the analysis of numerical results obtained from our program and their analytical results will be the aim of another paper.

\section{ Summary }
\noindent
Presented difference schemes and numerical algorithms allow to examine waves amplitudes evolution in non-stationary crystals with sufficiently precision. Numerical calculations show that even small non-stationary crystal deformations lead to considerable changes in the diffraction pattern. So, mathematical model and numerical method presented can be used in mathematical modeling of intensive beams passing through crystals.

\section{Acknowledgements }
\noindent
Author is pleased to thank Prof. V. N. Abrashin and Dr. A. O. Grubich for support and attention to work presented.

\vspace*{0.2truein}
\begin{center}
\hspace{10 cm} Svetlana Sytova 

\hspace{10 cm} Institute for Nuclear Problems

\hspace{10 cm} Belarus State University

\hspace{10 cm} Minsk 220050

\hspace{10 cm} Belarus

\hspace{10 cm} e-mail:sytova@inp.minsk.by

\end{center}

\end{document}